# Small Distance Increment Method for Measuring Complex Permittivity With mmWave Radar

Hang Song, Hyun Joon Kim, Mingxia Wan, Bo Wei, Takamaro Kikkawa, and Jun-ichi Takada

*Abstract*—Measuring the complex permittivity of material is essential in many scenarios such as quality check and component analysis. Generally, measurement methods for characterizing the material are based on the usage of vector network analyzer, which is large and not easy for on-site measurement, especially in high frequency range such as millimeter wave (mmWave). In addition, some measurement methods require the destruction of samples, which is not suitable for non-destructive inspection. In this work, a small distance increment (SDI) method is proposed to non-destructively measure the complex permittivity of material. In SDI, the transmitter and receiver are formed as the monostatic radar, which is facing towards the material under test (MUT). During the measurement, the distance between radar and MUT changes with small increments and the signals are recorded at each position. A mathematical model is formulated to depict the relationship among the complex permittivity, distance increment, and measured signals. By fitting the model, the complex permittivity of MUT is estimated. To implement and evaluate the proposed SDI method, a commercial off-the-shelf mmWave radar is utilized and the measurement system is developed. Then, the evaluation was carried out on the acrylic plate. With the proposed method, the estimated complex permittivity of acrylic plate shows good agreement with the literature values, demonstrating the efficacy of SDI method for characterizing the complex permittivity of material.

*Index Terms*—Complex permittivity measurement, small distance increment method, material characterization, mmWave radar, non-destructive inspection.

## I. INTRODUCTION

COMPLEX permittivity represents the characteristic of the material, which can be utilized to recognize different materials for various purposes such as security check [1]. The change of material component will influence the permittivity. Thus, measuring the permittivity can also be utilized for applications such as material deterioration, quality inspection and component analysis [2-3]. Furthermore, characterizing the materials can help construct the models of objects in the physical world. Therefore, it can facilitate the analysis of wireless communication channels [4].

Currently, there are several approaches to characterize the complex permittivity of the materials including the perturbation method [5], transmission line method [6], and free space method [7]. In the perturbation method, a resonant cavity is utilized and a small sample is inserted. By analyzing the resonant frequency shift caused by the material, the permittivity is estimated. In the transmission line method, a waveguide or transmission line is utilized and the materials are inserted into the waveguide. By measuring the reflection and transmission coefficients, the permittivity is estimated. Both the perturbation method and transmission line method require making small samples of the material which is not suitable if the material under test (MUT) cannot be destructed. Free space method is a contactless approach where the MUT is placed in free space. The antennas utilized to emit and receive the electromagnetic wave are placed at a certain distance from the MUT. The advantage of free space method is that the measurement can be conducted non-destructively. However, the above-mentioned methods generally require the vector network analyzer (VNA), which is large in high frequency range such as millimeter wave (mmWave). Therefore, it is not easy to be utilized for on-site measurement if the object is difficult to be moved.

For the purpose of accessibility and mobility, other signals such as WiFi and UWB are also utilized to characterize and identify different materials by analyzing the transmission signals through the MUT [8, 9]. However, it is difficult to achieve the transmission signals in some scenarios such as large and thick materials. This paper proposes a contactless and non-destructive method, small distance increment (SDI), for measuring the complex permittivity of material, where the transceiver is configured as monostatic radar and only reflection signal is utilized. This method can be deployed and integrated with off-the-self complex baseband radar modules. In SDI method, the radar is positioned perpendicular to the MUT. By changing the distance between radar and MUT gradually, the reflected signal from MUT is measured and recorded in complex value format. To estimate the complex permittivity of MUT, a mathematical model is proposed to establish the relationship between recorded signals and permittivity. Then, the fitting process is carried out with constraints to obtain the complex permittivity value.

Hang Song, Hyun Joon Kim, Mingxia Wan, and Junichi Takada are with the Department of Transdisciplinary Science and Engineering, Tokyo Institute of Technology, Tokyo 152-8550, Japan.

Bo Wei is with Department of Systems Innovation, School of Engineering, The University of Tokyo, Tokyo 113-8654, Japan, and also with Japan Science and Technology Agency (JST), PRESTO, Kawaguchi, Saitama 332-0012, Japan.

Takamaro Kikkawa is with the Research Institute for Nanodevice and Bio Systems, Hiroshima University, Hiroshima 739-8527, Japan.

In this work, a commercial off-the-shelf (COTS) mmWave radar module is utilized to evaluate the proposed method. Recently, the mmWave frequency modulated continuous wave (FMCW) radar has been widely exploited in autonomous driving [10-13], fall detection [14], vital signal monitoring [15-17], localization [18, 19] and other internet of things (IoT) applications [20-24]. Zhao *et al.* proposed an end-to-end human motion recognition module, CubeLearn which directly extracts features from raw radar signals [20]. Ren *et al.* proposed a processing pipeline to track different groups of people and realize the number counting within each group [21]. Gao *et al.* proposed a deep learning-based method to detect the open and concealed objects by using the mmWave radar [22]. Ninos *et al.* realized the gesture recognition in a real-time manner by proposing an efficient empirical feature extraction method and multi-layer perceptron networks [23]. Skaria *et al.* utilized three learning methods to classify different materials by examining the signatures of the radar reflection signals [24]. In contrast, this work integrates the COTS mmWave radar to deploy the SDI method and develop the system for measuring the complex permittivity of material.

In the measurement system, a stage with 5 phase stepper motor is employed to control the distance between radar and MUT. In experiment, the acrylic plate was used to evaluate the performance of the proposed method. A metal plate was utilized to calibrate the path loss between radar and MUT. After fitting the measured signals with the proposed model, the complex permittivity was estimated and the value showed good agreement with that in literature. The results demonstrate the efficacy of the proposed SDI method. It is also promising to implement SDI with other complex baseband modules for portable on-site permittivity measurement.

The remainder of this paper is organized as follows. Section II presents the principle of the proposed SDI method in detail. Section III explains the system development by using the FMCW radar and experimental setups for performance evaluation. Section IV presents the measurement results. Finally, conclusion is made in Section V.

II. PRINCIPLES OF SMALL DISTANCE INCREMENT METHOD

This section presents details of the principles of SDI method. Here, the plane wave is considered and it propagates perpendicularly onto the interface between MUT and air as shown in Fig. 1. The MUT is assumed to be a dielectric slab with the thickness of $d$.

A. *Theoretical formulation of SDI measurement*

Denote the first interface between air and MUT as $x = 0$. The relative permittivity and permeability of air and MUT are denoted as $\varepsilon_1$ and $\mu_1$, $\varepsilon_r$ and $\mu_r$, respectively. Here $\varepsilon_r$ is complex number as $\varepsilon_r' - j\varepsilon_r''$. For a general case, the other side of MUT is considered as another material with $\varepsilon_2$ and $\mu_2$.

Denote the electrical field of incident wave $E_i$ which propagates to the MUT as:

$$E_i = E_0 e^{-jk_1 x} \quad (1)$$

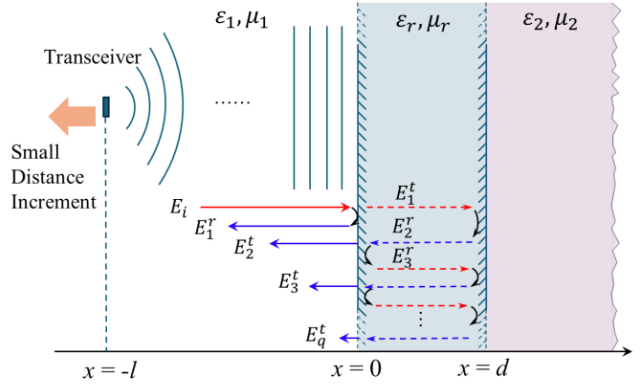

**Fig. 1.** Schematic diagram of the small distance increment method for measuring complex permittivity.

where $k_1$ is the wave number in the air, which is calculated by $\omega\sqrt{\varepsilon_1\varepsilon_0\mu_1}$. Here $\varepsilon_0$ is the permittivity of vacuum. When incident wave reaches the interface at $x = 0$, the first reflected wave $E_{r1}$ at interface at $x = 0$ is express as:

$$E_1^r = \Gamma_{1r} E_0 e^{+jk_1 x} \quad (2)$$

where $\Gamma_{1r}$ is the reflection coefficient at the interface between air and MUT. The subscript indicates the wave propagates from air to MUT.

The transmitted wave to MUT is expressed as:

$$E_1^t = T_{1r} E_0 e^{-jk_r x} \quad (3)$$

where $T_{1r}$ is the transmission coefficient at the interface between air and MUT. $k_r$ is the wave number in the MUT, which is calculated by $\omega\sqrt{\varepsilon_r\varepsilon_0\mu_r}$. Note that $k_r$ is complex number since $\varepsilon_r$ is complex permittivity. For most cases, the materials are non-ferromagnetic which means $\mu_1 = \mu_r$. Then, for the normal incidence case, $\Gamma_{1r}$ and $T_{1r}$ are calculated as:

$$\Gamma_{1r} = \frac{\sqrt{\varepsilon_1} - \sqrt{\varepsilon_r}}{\sqrt{\varepsilon_1} + \sqrt{\varepsilon_r}}, \quad T_{1r} = \frac{2\sqrt{\varepsilon_1}}{\sqrt{\varepsilon_1} + \sqrt{\varepsilon_r}} \quad (4)$$

After propagating through MUT, the wave is reflected at the interface between MUT and material 2. The reflected wave at $x = d$ can be expressed as:

$$E_2^r = \Gamma_{r2} E_1^t(x = d) e^{+jk_r(x-d)} = \Gamma_{r2} T_{1r} E_0 e^{-j2k_r d} e^{+jk_r x} \quad (5)$$

where $\Gamma_{r2}$ is the reflection coefficient at the interface between MUT and material 2. Then, the $E_2^r$ propagates back to the interface at $x = 0$. At this time, the transmitted wave can be expressed as:

$$E_2^t = T_{r1} E_2^r(x = 0) e^{-jk_1 x} = T_{r1} \Gamma_{r2} T_{1r} E_0 e^{-j2k_r d} e^{+jk_1 x} \quad (6)$$

where $T_{r1}$ is the transmission coefficient at the interface of $x = 0$ but the direction is from MUT to air. Meanwhile, the wave will be reflected partially at $x = 0$ as:

$$E_3^r = \Gamma_{r1} E_2^r(x = 0) e^{+jk_r x} = \Gamma_{r1} \Gamma_{r2} T_{1r} E_0 e^{-j2k_r d} e^{-jk_r x} \quad (7)$$

where $\Gamma_{r1}$ is also the reflection coefficient between air and MUT but the direction is reverse to $\Gamma_{1r}$.

Similar to the first transmitted wave $E_1^t$, $E_3^r$ will propagate through the MUT, be reflected at the interface of $x = d$, and bounce back to $x = 0$. Then, a new transmitted wave component will occur at $x = 0$ as:

$$E_3^t = T_{r1}\Gamma_{r2}T_{1r}E_0 e^{-j2k_r d}(\Gamma_{r1}\Gamma_{r2}e^{-j2k_r d})e^{+jk_1 x} \quad (8)$$

The multiple reflections will continue repeatedly at two interfaces and the general form of the transmitted wave can be expressed as:

$$E_q^t = T_{r1}\Gamma_{r2}T_{1r}E_0 e^{-j2k_r d}(\Gamma_{r1}\Gamma_{r2}e^{-j2k_r d})^{q-2}e^{+jk_1 x} \quad (9)$$

where $q$ indicates that the component is the $q$th transmitted wave at the interface of $x = 0$. Observing at $x = 0$, the total reflected wave will be the superposition of the first reflection $E_1^r$ and all the transmitted waves from MUT to the air. Therefore, the effective reflection coefficient at $x = 0$ can be calculated as:

$$\begin{aligned}\Gamma_{x=0} &= \frac{E_1^r + E_2^t + \cdots + E_q^t}{E_i} \\ &= \Gamma_{1r} + T_{r1}\Gamma_{r2}T_{1r}e^{-j2k_r d}(1 + \Gamma_{r1}\Gamma_{r2}e^{-j2k_r d} + \\ & \quad \ldots + (\Gamma_{r1}\Gamma_{r2}e^{-j2k_r d})^{q-2}) \\ &= \Gamma_{1r} + T_{r1}\Gamma_{r2}T_{1r}e^{-j2k_r d}\frac{1 - (\Gamma_{r1}\Gamma_{r2}e^{-j2k_r d})^{q-1}}{1 - \Gamma_{r1}\Gamma_{r2}e^{-j2k_r d}} \\ &= \Gamma_{1r} + \frac{T_{r1}\Gamma_{r2}T_{1r}e^{-j2k_r d}}{1 - \Gamma_{r1}\Gamma_{r2}e^{-j2k_r d}} \quad (q \to \infty)\end{aligned} \quad (10)$$

In SDI, the measurement is carried out in monostatic configuration where the transceiver is placed at $x = -l$. Therefore, the measured reflection coefficient at the transceiver position can be written as:

$$\Gamma_{x=-l} = e^{j2k_1 l}\Gamma_{x=0} \quad (11)$$

Generally, if $\varepsilon_1$, $\varepsilon_2$, $d$, and $l$ are known, $\varepsilon_r$ can be calculated by $\Gamma_{x=-l}$ using (11). In high frequency range, since the wave number is large, small measurement deviation in $l$ or $d$ may lead to large error in the estimation result. Therefore, in SDI, the measurement is conducted $M$ times considering measurement error. In each measurement, the distance between transceiver and MUT is changed with a small increment $\Delta l$. Then, the $\varepsilon_r$ is estimated by minimizing the following function:

$$\min_{\varepsilon_r} \sum_{m=0}^{M-1}\left\|\Gamma_{x=-l-m\Delta l} - e^{j2k_1(l+m\Delta l)}\Gamma_{x=0}\right\|^2 \quad (12)$$

In this work, the FMCW radar is utilized to implement SDI in a practical way. In FMCW radar, the emitted signal can be expressed as:

$$S_T = A_0 \cos(2\pi f_0 t + \pi \frac{B}{T_c}t^2 + \varphi_0) \quad (13)$$

where $A_0$ is the amplitude of the signal. $f_0$ and $\varphi_0$ are the start frequency and initial phase of the FMCW signal. $B$ and $T_c$ are the bandwidth and chirp duration. When the emitted signal encounters object, reflection occurs and the reflected signals can be written as:

$$s_R = A_0 \Gamma L_{path} \cos(2\pi f_0(t-\tau) + \pi \frac{B}{T_c}(t-\tau)^2 + \varphi_0) \quad (14)$$

where $\Gamma$ is the reflection coefficient of the object. $\tau$ is the time delay of the reflected wave. $L_{path}$ is the path loss between the radar and object. For plane wave case, path loss can be estimated using $e^{j2k_1 l}$ in (11). However, the rigorous plane wave condition is difficult to be achieved in the whole path between transceiver and object. Here, the general term $L_{path}$ is used in practical case. Then, $S_T$ and $S_R$ is mixed and the low-pass filter is applied to obtain the intermediate frequency (IF) signal as:

$$s_{IF} = LFP(S_T \cdot S_R) = \frac{A_0^2}{2}\Gamma L_{path}\cos(2\pi\frac{B}{T_c}\tau \cdot t + 2\pi f_0 \tau) \quad (15)$$

Then, $S_{IF}$ is digitized and recorded. Rewrite $S_{IF}$ as discrete form in complex format:

$$s_{IF}[n] = \frac{A_0^2}{2}\Gamma L_{path} e^{j2\pi(\frac{B}{T_c}\tau n\Delta t + f_0 \tau)} \quad (16)$$

where $\Delta t$ is the sampling interval in time domain. Applying the discrete Fourier transform to $s_{IF}[n]$, the IF signal can be expressed in frequency domain as:

$$\begin{aligned}S_{IF}[k] &= \sum_{n=0}^{N-1} s_{IF}[n] \cdot e^{-j2\pi \frac{nk}{N}} \\ &= \sum_{n=0}^{N} \frac{A_0^2}{2}\Gamma L_{path} e^{j2\pi(\frac{B}{T_c}\tau n\Delta t + f_0 \tau)} \cdot e^{-j2\pi\frac{nk}{N}} \\ &= \frac{A_0^2}{2}\Gamma L_{path} e^{j2\pi f_0 \tau} \sum_{n=0}^{N} e^{j\frac{2\pi n}{N}(\frac{B}{T_c}\tau N\Delta t - k)}\end{aligned} \quad (17)$$

where $N$ is the total sample number. By setting the distance between radar and MUT larger than the far field condition, $\Gamma$ in (17) can be replaced by $\Gamma_{x=0}$ since the wave in far field can be regarded as local plane wave. If the thickness of MUT $d$ is larger than the range resolution of the radar, the multiple reflections within the MUT will not be simply summed up since the frequency is changed. After Fourier transform, the reflections will be shown as different peaks in $S_{IF}[k]$. The frequency with the maximum amplitude corresponds to the first reflection $\Gamma_{1r}$. In this work, the condition when $d$ is large is considered. Since $L_{path}$, $A_0$ are unknown and difficult to be precisely measured, it is not possible to estimate the effective reflection coefficient directly from the measured signal. Therefore, another calibration measurement is carried out with metal plate when the plate is set at the same position of MUT. Note that the reflection coefficient of metal is known as -1 and the path loss is the same. Thus, the received signal in calibration can be expressed as:

$$S_{IF}^{metal}[k] = (-1) \cdot \frac{A_0^2}{2} L_{path} e^{j2\pi f_0 \tau} \sum_{n=0}^{N} e^{j\frac{2\pi n}{N}(\frac{B}{T_c}\tau N\Delta t - k)} \quad (18)$$

Then, the measured reflection coefficient by using the calibration can be expressed as:

$$\Gamma_{mea} = \frac{S_{IF}[k_{max}]}{S_{IF}^{metal}[k_{max}]} \quad (19)$$

$$= \frac{\frac{A_0^2}{2}\Gamma_{1r}L_{path}e^{j2\pi f_0 \tau_1}\sum_{n=0}^{N}e^{j\frac{2\pi n}{N}(\frac{B}{T_c}\tau_1 N\Delta t - k)}}{(-1)\cdot\frac{A_0^2}{2}L_{path}e^{j2\pi f_0 \tau_2}\sum_{n=0}^{N}e^{j\frac{2\pi n}{N}(\frac{B}{T_c}\tau_2 N\Delta t - k)}}$$

where $k_{max}$ is the bin of frequency with the maximum amplitude. Here, the time delays from MUT and metal are denoted by different terms $\tau_1$ and $\tau_2$ which are ideally the same. Then, $\Gamma_{1r}$ equals $-S_{IF}[k_{max}]/S_{IF}^{metal}[k_{max}]$. However, in the real condition, $\tau_1$ and $\tau_2$ might have small difference due to the error in distance $l$. Since $f_0$ is large, the small difference $(\tau_1 - \tau_2)$ can introduce non-negligible phase difference. While the difference is negligible in the summation term in (18) since $\frac{B}{T_c}N\Delta t$ is small. Meanwhile, there may be other factors which cause a phase difference $\Delta\varphi_0$ between the MUT measurement and calibration. Therefore, the measured reflection coefficient can be expressed as:

$$\Gamma_{mea} = \Gamma_{1r} e^{j2\pi f_0(\tau_1-\tau_2)} e^{j\Delta\varphi_0} = \frac{1-\sqrt{\varepsilon_r}}{1+\sqrt{\varepsilon_r}} e^{j\Delta\varphi} \quad (20)$$

where $\Delta\varphi$ is a general phase term which includes all factors. Here, the relative permittivity of air is considered as 1. Since $\Delta\varphi$ is unknow, it is not possible to estimate $\varepsilon_r$ with only one measurement. Therefore, the proposed SDI method is applied. During calibration, the distance between the radar and metal is changed with a small increment $\Delta l$. Since the increment is very small, the path loss can be considered not changed. But the introduced phase rotation is non-negligible. Therefore, the measured reflection coefficient by using the $m$th calibration measurement data can be expressed as:

$$\Gamma_{mea}(m) = \frac{1-\sqrt{\varepsilon_r}}{1+\sqrt{\varepsilon_r}} e^{j\Delta\varphi} e^{-j2\pi f_0 \frac{2m\Delta l}{v_c}} \quad (21)$$

where $v_c$ is the speed of light. Then, the $\varepsilon_r$ is estimated by minimizing the following function:

$$\min_{\varepsilon_r} \sum_{m=0}^{M-1} \left\| \Gamma_{mea}(m) - \frac{1-\sqrt{\varepsilon_r}}{1+\sqrt{\varepsilon_r}} e^{j\Delta\varphi} e^{-j2\pi f_0 \frac{2m\Delta l}{v_c}} \right\|^2 \quad (22)$$

where $M$ is the number of increment times which is the same as the measurement times in (12).

### B. Fitting procedure for $\varepsilon_r$

$\varepsilon_r$ is complex number and the values of real part and imaginary part have boundaries to make it physically meaningful. Therefore, (22) is reformulated into real-value format for fitting with bound constraints. For conciseness, denote $\varepsilon_r$ and $\Delta\varphi$ as $a - jb$ and $c$, respectively. Replace $2\pi f_0 \frac{2\Delta l}{v_c}$ as $C_1$ which is a constant value. Then, the right side of (20) is rewritten as:

$$\frac{1-\sqrt{a-jb}}{1+\sqrt{a-jb}} e^{j(c-C_1 m)} \quad (23)$$

where there are three unknown variables, $a$, $b$, and $c$. $\sqrt{a-jb}$ can be written as:

$$\sqrt{a-jb} = \frac{\sqrt{2}}{2}\left(\sqrt{\sqrt{a^2+b^2}+a} - j\sqrt{\sqrt{a^2+b^2}-a}\right) \quad (24)$$

Substitute (23) into (22), thus (22) can be further expressed as:

$$\frac{1-\frac{\sqrt{2}}{2}\left(\sqrt{\sqrt{a^2+b^2}+a}-j\sqrt{\sqrt{a^2+b^2}-a}\right)}{1+\frac{\sqrt{2}}{2}\left(\sqrt{\sqrt{a^2+b^2}+a}-j\sqrt{\sqrt{a^2+b^2}-a}\right)} e^{j(c-C_1 m)} \quad (25)$$

$$= \frac{1-\sqrt{a^2+b^2}+j\sqrt{2}\sqrt{\sqrt{a^2+b^2}-a}}{1+\sqrt{a^2+b^2}+\sqrt{2}\sqrt{\sqrt{a^2+b^2}+a}} \cdot (\cos(c-C_1 m) + j\sin(c-C_1 m))$$

$$= \frac{1}{g_1}(g_2 + jg_3)$$

where real-valued terms $g_1$, $g_2$, and $g_3$ are written as follows:

$$g_1 = 1 + \sqrt{a^2+b^2} + \sqrt{2}\sqrt{\sqrt{a^2+b^2}+a} \quad (26)$$

$$g_2 = \left(1-\sqrt{a^2+b^2}\right)\cdot \cos(c-C_1 m) \quad (27)$$
$$-\sqrt{2}\sqrt{\sqrt{a^2+b^2}-a}\cdot \sin(c-C_1 m)$$

$$g_3 = \left(1-\sqrt{a^2+b^2}\right)\cdot \sin(c-C_1 m) \quad (28)$$
$$+\sqrt{\sqrt{a^2+b^2}-a}\cdot \cos(c-C_1 m)$$

Substitute (25) into (22), the minimizing problem is reformulated into a real-valued format as:

$$\min_{a,b,c} \sum_{m=0}^{M-1} \left( \left\| \text{Re}(\Gamma_{mea}(m)) - \frac{g_2}{g_1} \right\|^2 + \left\| \text{Im}(\Gamma_{mea}(m)) - \frac{g_3}{g_1} \right\|^2 \right) \quad (29)$$

where the constraints of $a$ and $b$ are based on the fact that the real part of complex permittivity should be larger than 1 and the imaginary part should be larger than 0. $c$ is limited to one phase cycle. The relationship between the measured reflection coefficient and the complex permittivity are nonlinear. Finally, a nonlinear least-squares algorithm for minimization subject to bounds is utilized to fit the permittivity [25].

### III. MEASUREMENT SYSTEM AND EXPERIMENT SETUP

To evaluate the proposed SDI method with FMCW radar, the experiment system is developed. Meanwhile the experiment setup and procedure are presented in this section.

### A. Measurement system development

The structure of the developed system is shown in Fig. 2. The system mainly consists of radar module, one-axis manual stage, motorized linear stage, rotation stage, digimatic micrometer, and laser. The FMCW radar (T14, S-Takaya) is a COTS module which is based on the TI-IWR1443 chipset. The center frequency is 79 GHz, respectively. The radar is mounted on a customized holder and the holder is installed on a rotation stage (KSP-606M, SigmaKoki). A laser (VLM-635-63-LPO-200, Quarton) is also mounted on the same holder with the radar for adjusting the direction of radar to be perpendicular to the MUT and metal. To realize the small distance increment, a motorized linear stage (OSMS20-35(X), SigmaKoki) equipped with a 5-phase stepping motor is utilized. The lead of the ball screw is 1 mm and the resolution of the stage can reach 1 $\mu$m in half-step drive mode, which can enable precise control of position step during the measurement. On the linear stage, a one-axis manual stage (TSD-601C, SigmaKoki) and rotation stage are mounted. Facing to the radar, the metal for calibration is mounted on a metal holder and the holder is installed on the cascaded stages, which enables the adjustment of the position and angle. Since in the proposed SDI method, the distances from radar to MUT and to the metal are assumed to be the same, a digimatic micrometer (ID-SX2, Mitutoyo) is utilized to record the positions for ensuring this condition. The linear stage, micrometer, and rotation stage are all fixed on an optical baseplate at the same line in the measurement system.

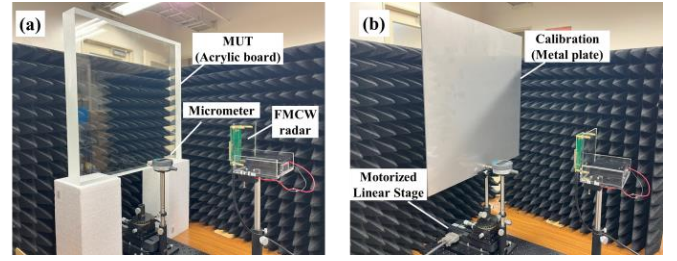

**Fig. 3.** Measurement setup of (a) MUT and (b) calibration.

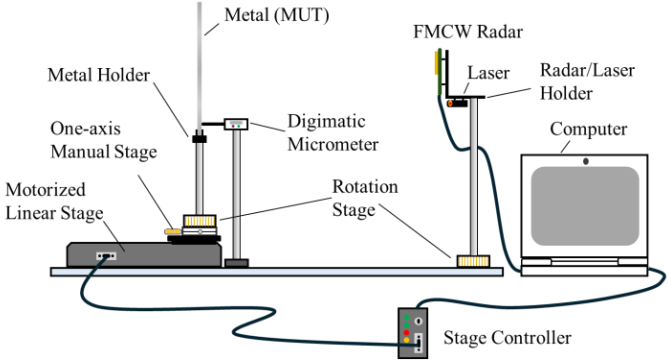

**Fig. 2.** Measurement system structure of SDI.

*B. Experiment setup and procedure*

During experiment, the MUT is firstly measured and then the metal plate is measured for calibration. In this work, the acrylic board is utilized as MUT for evaluation. As shown in Fig. 3(a), the MUT is placed on the Styrofoam. The needle of digimatic micrometer is contacted with the MUT. Then, the laser is turned on and pointed to the MUT. The rotation stage which holds the radar/laser module is adjusted. The purpose of the adjustment is to make the reflection of laser point back to the source, which ensures that the module is perpendicular to MUT. After adjustment, the measurement is carried out with FMCW radar and the data are recorded into the computer. The readout of micrometer is memorized to record the position of MUT.

After being measured, the MUT is moved away and the metal plate is installed for calibration as shown in Fig. 3(b). The positions of the whole system including the micrometer and radar are not changed during the replacement. After the installation of metal plate, the one-axis manual stage which holds the metal is adjusted to make the micrometer show the same value as that of MUT, ensuring that the metal is set at the same position as that of the MUT. Meanwhile, the rotation stage which holds the metal is also adjusted to make sure that laser reflection points back to the source, satisfying the perpendicular condition. After the adjustment, the calibration step is carried out. During the calibration, the motorized linear stage and radar module are controlled simultaneously by the computer. The metal is moved backward with a constant step of $\Delta l$ and the measurement is conducted. This operation is repeated for $M$ times. And the data is recorded each time after the metal is moved.

By completing all the measurement procedures, the recorded data are post-processed for estimating the complex permittivity with the proposed method depicted in Section II.

## IV. MEASUREMENT RESULTS

Before measuring complex permittivity, the performance of radar is evaluated by using metal plate. Since the calculation of permittivity used in this work is based on the plane wave assumption at MUT boundary, the measurement data should satisfy this condition for post-processing. Then, the estimation of MUT complex permittivity is carried out using the proposed method.

*A. Far field: amplitude and phase change with SDI*

When implementing SDI with FMCW radar, the condition of far field is considered when deriving the measurement principles. Therefore, in the experiment, the radar needs to be placed at an appropriate distance. The limit between the near and far field is determined by Fraunhofer distance:

$$d_F = \frac{2D^2}{\lambda} \quad (30)$$

where $D$ is the length of antenna and $\lambda$ is the wavelength. When the distance is larger enough than $d_F$, it is considered as the far field region. The length of antenna in the radar module is around 1.5 cm and the wavelength is about 3.8 mm. Therefore, $d_F$ is around 11.8 cm. To satisfy the far field condition, MUT is expected to be set far from 11.8 cm. On the other hand, if the radar is set too far away, it requires extra spaces to conduct the measurement which may result in difficulty in the practical on-site measurement. In this work, the distance is chosen as 25 cm which is about two times of the Fraunhofer distance. To verify the far field condition, the metal is utilized and the radar is set at the distance of 25 cm to the metal. Then the measurement is carried out with the procedure depicted in Section III.B. The distance increment $\Delta l$ is 0.1 mm and the number of total movement times is 40.

Figure 4(a) and (b) show the amplitudes and phases of received signal $S_{\text{IF}}^{\text{metal}}[k_{max}]$ along time. Each plot in different color represents the result at a certain distance. It can be observed that the standard deviation of amplitude with the change of time is from 7 to 12, which is equivalent to around 0.03% to 0.05% of the average amplitude. Here the unit amplitude is arbitrary unit since it is digitized. This shows the temporal stability of the measurement. Regarding the phase, the standard deviation is from 0.2° to 0.8° along time which is also stable. Fig. 4(c) and (d) shows the amplitudes and phases at different measurement points during the movement. It can be observed that the amplitude is slowly decreasing while oscillating along the distance increment. Take the mean value along time axis for each measurement point. The maximum difference among all measurement points is 310, which is about 1.22% of the average amplitude through all the measurements. Therefore, $L_{path}$ in (17) can be considered unchanged during the measurement. Regarding the phase, it can be observed in Fig. 4(d) that phase is linearly changed along the movement. Take the mean value of phase along time axis for each measurement point. Then unwrap the phase and transform radian into degree. By fitting the phase change with linear regression, the regression coefficient is -187.9°/mm and the coefficient of determination $R^2$ is 1, which demonstrates the rigorous linearity of the phase change with distance increment. Theoretically, if the wave front is planar, the increment of distance by 0.1 mm, which is 0.2 mm for round-trip, will cause the phase shift of 18.9° since the wavelength is around 3.8 mm. Compared with the measured data, the phase shows good agreement with the theoretical value. Therefore, it can be considered that the far field condition is satisfied at the distance of 25 cm.

### B. Change of reflection coefficient in SDI

When the MUT or metal is in the far field region of the radar, the plane wave can be considered at the interface between air and MUT or metal. Meanwhile assume that the measurement conditions between MUT and metal are the same where the initial phase term $\Delta\varphi$ in (20) is 0. If the complex permittivity is known, the measured reflection coefficients by SDI method can be calculated by (20). Consider three arbitrarily generated materials with the permittivity of $2 - j0.1$, $3 - j0.15$ and $7 - j0.3$. Note that these values are not necessarily corresponding to any real material. The theoretical reflection coefficients with regards to the distance increment of 0.1 mm are plotted in Fig. 5. From the plot of the real part of reflection coefficient during SDI, it can be found that the value repeats periodically along with distance increment. The same behavior can be also observed in the imaginary part of reflection coefficient, but the phase is different from the real part. When comparing the real part of reflection coefficient among three materials, it can be observed that the maximum values are different. Basically, the larger permittivity results in larger reflection. The measurement result from MUT is also included in Fig. 5. The result for MUT is calculated by (18) with the SDI method and the measured data of MUT and calibration data of metal are utilized. From the result, the periodical change of the reflection coefficient is also observed in the measurement data. While a phase shift is significant at $\Delta l = 0$, compared with the plots of ideal cases.

The amplitude and phase of the measured reflection coefficient $\Gamma_{\text{mea}}(m)$ with SDI method are shown in Fig. 6. Since $\Gamma_{\text{mea}}(m)$ is calculated by dividing the MUT data by the calibration data, the amplitude of $\Gamma_{\text{mea}}(m)$ is inversely proportional to the amplitude of $S_{\text{IF}}^{\text{metal}}[k_{max}]$ compared with Fig. 4(c). Meanwhile, the phase is also linearly changing along with the distance increment. By fitting the measured reflection coefficient data, the permittivity is estimated.

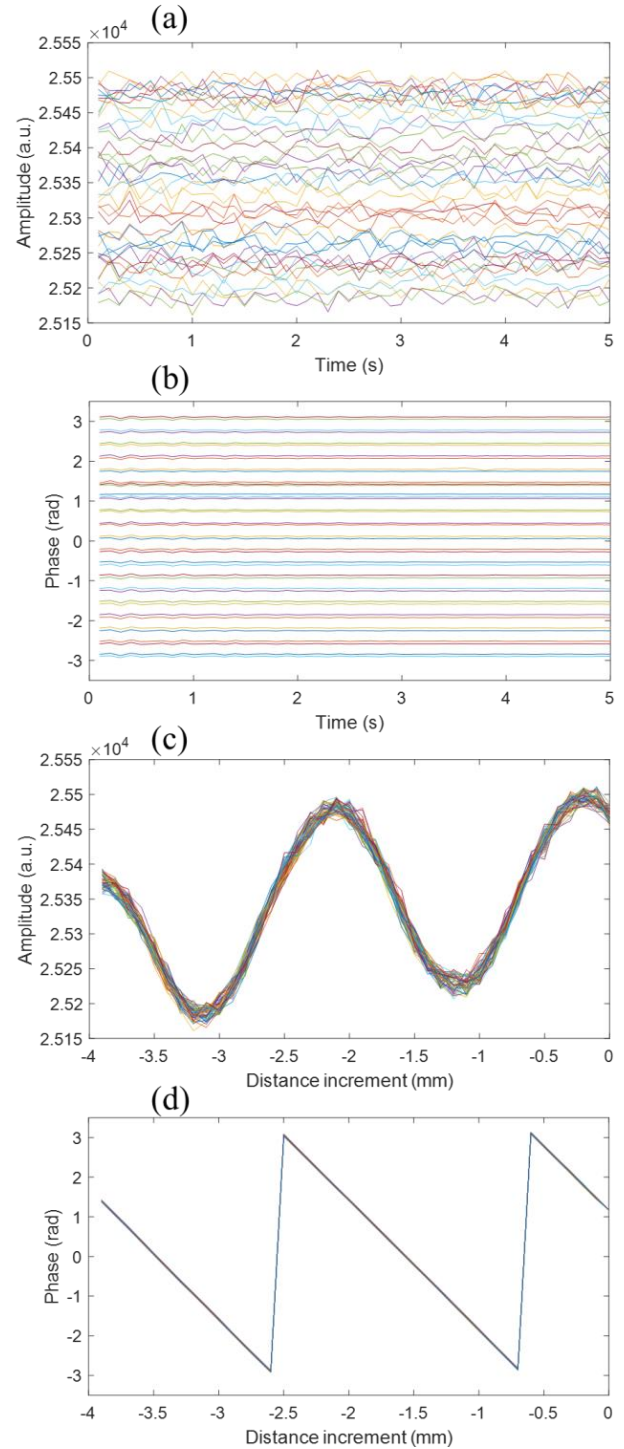

**Fig. 4.** Received IF signals at the frequency with maximum amplitude after Fourier transform. (a) Amplitude change along time. (b) Phase change along time. Each plot in the figure corresponds to the data for a certain measurement point. (c) Amplitude change along distance increment. (d) Phase change along distance increment. Each plot in the figure corresponds to the data for a certain sampling time.

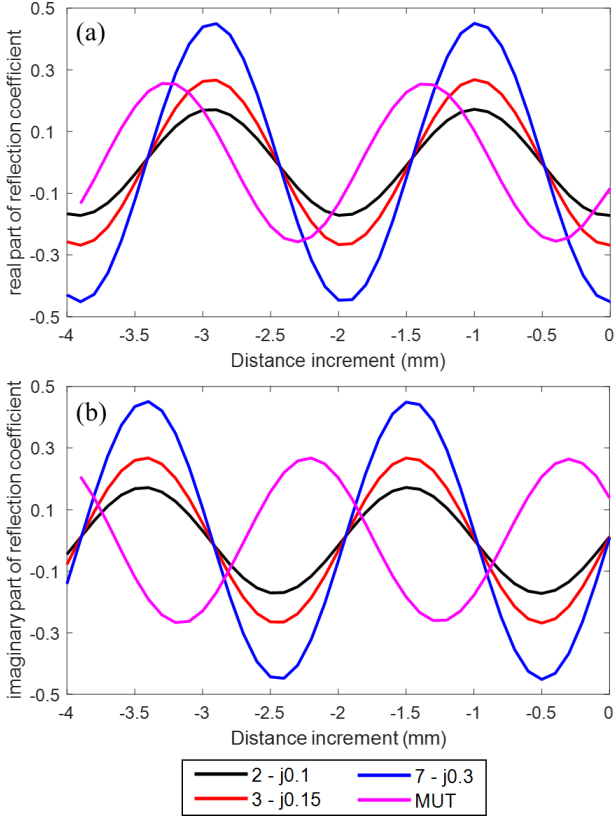

**Fig. 5.** Reflection coefficients during SDI with arbitrary materials and the measurement result of MUT. (a) Real part. (b) Imaginary part.

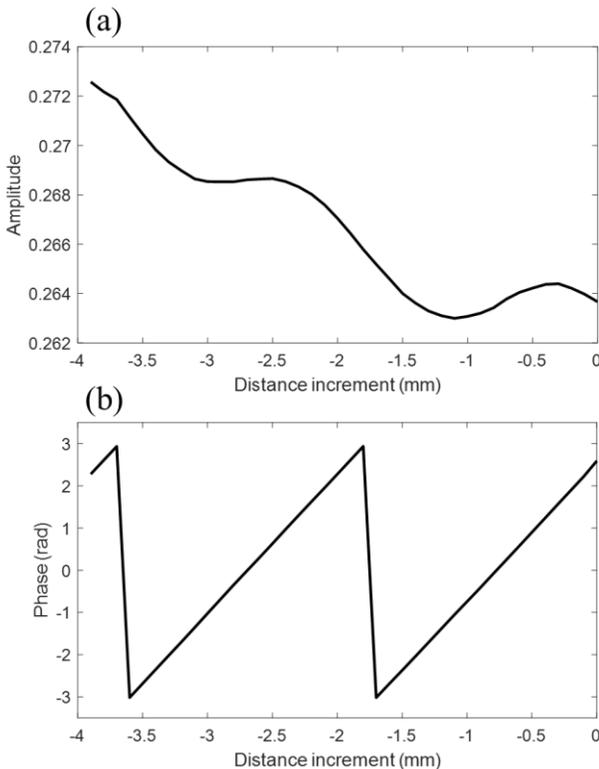

**Fig. 6.** Measured reflection coefficients of MUT with SDI. (a) Amplitude. (b) Phase.

*C. Estimation of the complex permittivity of MUT*

Using the measured reflection coefficient data shown in Fig. 6, the fitting process for the complex permittivity is carried out by formulating the minimization problem as (28). After fitting, the estimated complex permittivity of acrylic plate is $2.60 - j0.1$. Using the fitted complex permittivity, the fitted reflection coefficients along with the distance increment can be calculated by (21). The measured and fitted reflection coefficients are shown in Fig. 7. It can be observed that both the real and imaginary part of the reflection coefficient are fitted well. From the literature [26-29], the real part of permittivity near 79 GHz is in the range of 2.53~2.65 while the imaginary part is in the range of 0.02~0.05. The good agreement between the measurement result and the literature values demonstrates the efficacy of the proposed SDI method with FMCW radar.

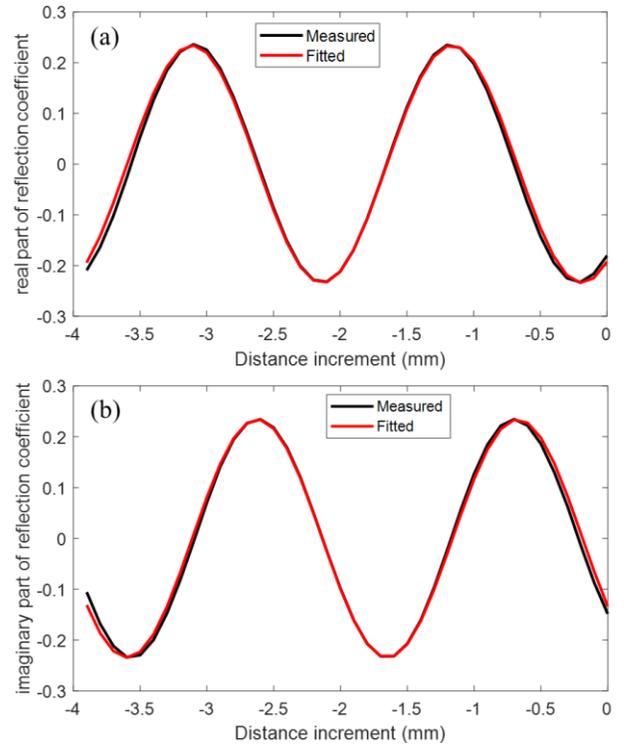

**Fig. 7.** Measured reflection coefficients of MUT and the fitted results. (a) Amplitude. (b) Phase.

V. CONCLUSION

This paper proposed SDI, a small distance increment method for contactless onsite measurement of the complex permittivity of material non-destructively. The proposed method can be deployed with any complex baseband radar modules. In this work, FMCW radar module was utilized to realize and implement the SDI method. In SDI, the measurement was conducted by changing the distance between transceiver and materials in small distance increment step. Meanwhile, the metal plate was utilized as calibration. A mathematical model was proposed to establish the relationship between SDI measurement data on MUT and the complex permittivity. To verify the feasibility, the measurement on an acrylic board was carried out. By fitting the measurement data, the estimated complex permittivity showed good agreement with the values in literatures. With the proposed method, it is promising to realize on-site complex permittivity measurement with different compact radar modules for a wide range of applications.